\documentclass[twocolumn,showpacs,preprintnumbers]{revtex4}

\begin{document}

\title{Solution of polynomiality and positivity constraints on generalized
parton distributions}
\author{P.~V.~Pobylitsa}
\affiliation{Institute for Theoretical Physics II, Ruhr University Bochum, D-44780
Bochum, Germany\\
and Petersburg Nuclear Physics Institute, Gatchina, St. Petersburg, 188300,
Russia}

\begin{abstract}
An integral representation for generalized parton distributions
of spin-0 hadrons is suggested,
which satisfies both polynomiality and positivity constraints.
\end{abstract}

\maketitle

\section{Introduction}

Generalized parton distributions (GPDs) \cite%
{MRGDH-94,Radyushkin-96-a,Radyushkin-96,Ji-97,Ji-97-b,CFS-97,Radyushkin-97,Radyushkin-review,GPV,BMK-2001}
appear in the context of the QCD factorization in various hard exclusive
phenomena including deeply virtual Compton scattering and hard exclusive
meson production. Among several general constraints on GPDs an important
role is played by the polynomiality of the Mellin moments \cite{Ji-97-b} and
by the positivity bounds \cite{Martin-98,
Radyushkin-99,PST-99,Ji-98,DFJK-00,Burkardt-01,Pobylitsa-01,Pobylitsa-02,Diehl-02,Burkardt-02-a,Burkardt-02-b,Pobylitsa-02-c}.
In this paper we suggest a
representation for GPDs which automatically satisfies both positivity and
polynomiality constraints. The analysis is restricted to the case
of spin-0 hadrons (e.g. pions) but various types of partons will be covered.

We use the following definition of GPDs: 
\begin{equation}
H^{(N)}(x,\xi ,t)=\int \frac{d\lambda }{2\pi }\exp (i\lambda x)\langle
P_{2}|O^{(N)}(\lambda ,n)|P_{1}\rangle \,.  \label{F-N-def}
\end{equation}%
Here $|P_{k}\rangle $ is the hadron state with momentum $P_{k}$. The
light-like vector $n$,
\[
n^{2}=0\,,
\]
is normalized by the condition
\begin{equation}
n(P_{1}+P_{2})=2\,.
\end{equation}
We use the standard notation of Ji \cite{Ji-98} for the parameters $\Delta $, $%
t $ and $\xi $%
\begin{equation}
\Delta =P_{2}-P_{1}\,,\quad \xi =-\frac{1}{2}(n\Delta ),\quad t=\Delta
^{2}\,.  \label{Ji-variables}
\end{equation}%
The definitions of light-ray operators $O^{(N)}(\lambda ,n)$ for various
types of partons are listed in the Table \ref{table-N}. 
\begin{table}[tbp]
\begin{tabular}{||l|l|l||}
\hline\hline
Parton\vphantom{$\int\limits_0^1$} & $O^{(N)}(\lambda ,n)$ & $N$ \\ 
\hline\hline
scalar\vphantom{$\int\limits_0^1$} & $\phi ^{\dagger }\left( -\frac{\lambda n%
}{2}\right) \phi \left( \frac{\lambda n}{2}\right) $ & $0$ \\ \hline
quark\vphantom{$\int\limits_0^1$} & $\frac12\bar{\psi}\left( -\frac{\lambda n}{2}%
\right) (n\cdot \gamma )\psi \left( \frac{\lambda n}{2}\right) $ & $1$ \\ 
\hline
gluon\vphantom{$\int\limits_0^1$} & $n^{\mu }G_{\mu \nu }^{a}\left( -\frac{%
\lambda n}{2}\right) n_{\rho }G^{a,\nu\rho}\left( \frac{\lambda n}{2}\right) 
$ & $2$ \\ \hline\hline
\end{tabular}
\label{table-N} 
\caption{Light-ray operators $O^{(N)}(\lambda ,n)$ for various types of partons
and the corresponding parameter $N$.}
\end{table}
We have included the scalar field $\phi $ into this table since the
positivity bounds are more general than their applications in QCD. The last
column of this table contains the number $N$ of factors $n^{\mu }$ appearing
in the light-ray operator $O^{(N)}(\lambda ,n)$. This number $N$ plays an
important role in the formulation of the positivity bounds and of the
polynomiality conditions and we include $N$ in the notation (\ref{F-N-def})
of GPD $H^{(N)}(x,\xi ,t)$.

The structure of the paper is as follows. Section
\ref{polynomiality-positivity-section} contains a brief description of the
polynomiality and positivity constraints on GPDs. In Section \ref%
{modified-DD-section} we introduce a modified version of the double
distribution representation for GPDs which slightly differs from the
standard one but is more relevant for our aims. In Section \ref%
{DD-ansatz-section} an ansatz for the double distribution is suggested. In
the remaining part of the paper we show that this ansatz leads to GPDs which
obey both polynomiality and positivity constraints. In Section \ref%
{GPDs-section} we compute the GPDs corresponding to our ansatz for double
distributions. Section \ref{forward-section} contains the check of the
positivity of the forward parton distributions. The positivity bounds on GPDs are
verified in Section \ref{positivity-check-section}. Appendix \ref%
{r-k-appendix} contains the derivation of the general solution of the
positivity bounds (without the polynomiality constraints).

\section{Polynomiality and positivity}

\label{polynomiality-positivity-section}

Whatever limited our knowledge about GPDs is, there are two basic
constraints: polynomiality and positivity. The polynomiality means that
Mellin moments in $x$ of GPD $H^{(N)}(x,\xi ,t)$%
\begin{equation}
\int\limits_{-1}^{1}dx\,x^{m}H^{(N)}(x,\xi ,t)=P_{m+N}(\xi ,t)
\label{polynomiality-power}
\end{equation}%
must be polynomials in $\xi $ of degree $m+N$.

Various inequalities for GPDs suggested in the Refs. \cite{Martin-98,
Radyushkin-99,PST-99,Ji-98,DFJK-00,Burkardt-01,Pobylitsa-01,Pobylitsa-02,Diehl-02,Burkardt-02-a,Burkardt-02-b}
can be considered as particular cases of the general positivity bound on
GPDs derived in Ref. \cite{Pobylitsa-02-c}. This general positivity bound has
a relatively simple formulation in the impact parameter representation \cite%
{Burkardt-01,Diehl-02,Burkardt-02-a,Burkardt-02-b,Burkardt-00}. The impact parameter $%
b^{\perp }$ appears via the Fourier transformation of the $\Delta ^{\perp }$
dependence of GPDs.
If the transverse plane is orthogonal to vectors $n$ and $P_1+P_2$, then
the transverse component $\Delta ^{\perp }$ of the
momentum transfer $\Delta $ (\ref{Ji-variables}) is connected with the
variable $t=\Delta ^{2}$ by the relation%
\begin{equation}
t=-\frac{|\Delta ^{\perp }|^{2}+4\xi ^{2}M^{2}}{1-\xi ^{2}}\,.
\end{equation}%
Here $M$ is the mass of the hadron. We define the GPD in the impact
parameter representation as follows:
\[
\tilde{F}^{(N)}\left( x,\xi ,b^{\perp }\right) =\int \frac{d^{2}\Delta
^{\perp }}{(2\pi )^{2}}\exp \left[ i(\Delta ^{\perp }b^{\perp })\right]
\]
\begin{equation}
\times H^{(N)}\left( x,\xi ,-\frac{|\Delta ^{\perp }|^{2}+4\xi ^{2}M^{2}}{%
1-\xi ^{2}}\right) \,.  \label{F-impact-def}
\end{equation}%
Here notation $\tilde{F}^{(N)}$ is used in order to avoid confusion with the
nucleon GPD $\tilde{H}$ and to keep the compatibility with the notation of
Ref. \cite{Pobylitsa-02-c}, where the following inequality was derived:

\[
\,\int\limits_{-1}^{1}d\xi \int\limits_{|\xi |}^{1}dx(1-x)^{-N-4}p^{\ast
}\left( \frac{1-x}{1-\xi }\right) p\left( \frac{1-x}{1+\xi }\right)
\]
\begin{equation}
\times \tilde{F}^{(N)}\left( x,\xi ,\frac{1-x}{1-\xi ^{2}}b^{\perp }\right)
\geq 0\,.  \label{ineq-q}
\end{equation}%
This inequality was obtained in Ref. \cite{Pobylitsa-02-c} for the case $N=1$
and the generalization to arbitrary $N$ is straightforward.

Inequality (\ref{ineq-q}) should hold for any function $p(z)$. Therefore we
actually deal with an infinite set of positivity bounds on the GPD. The
general inequality (\ref{ineq-q}) covers various inequalities suggested for
GPDs \cite{Martin-98,
Radyushkin-99,PST-99,Ji-98,DFJK-00,Burkardt-01,Pobylitsa-01,Pobylitsa-02,Diehl-02,Burkardt-02-a,Burkardt-02-b}
as particular cases with some special choice of functions $p(z)$.

It is well known that the double distribution representation \cite%
{MRGDH-94,Radyushkin-96-a,Radyushkin-97} with the $D$-term \cite{PW-99}%
\[
H(x,\xi ,t)=\int\limits_{|\alpha |+|\beta |\leq 1}d\alpha d\beta \delta
(x-\xi \alpha -\beta )\bar{F}_{D}(\alpha ,\beta ,t)
\]
\begin{equation}
+\theta (|\xi |-|x|)D\left( \frac{x}{\xi },t\right) \mathrm{sign}(\xi )
\label{DD-representation}
\end{equation}%
guarantees the polynomiality property (\ref{polynomiality-power}). Another
interesting parametrization for GPDs supporting the polynomiality was
suggested in Ref. \cite{PS-02}.

The positivity bound on GPDs (\ref{ineq-q})  is equivalent
to the following representation for GPDs in the impact parameter
representation (see Appendix \ref{r-k-appendix}) in the region $x>|\xi|$:
\[
\tilde{F}^{(N)}\left( x,\xi ,b^{\perp }\right) =(1-x)^{N+1}
\]
\begin{equation}
\times \sum\limits_{n}Q_{n}\left( \frac{1-x}{1+\xi },(1-\xi )b^{\perp
}\right) Q_{n}\left( \frac{1-x}{1-\xi },(1+\xi )b^{\perp }\right)
\label{pos-representation}
\end{equation}%
with arbitrary real functions $Q_{n}$. Instead of the discrete summation over $n$
one can use the integration over continuous parameters.

Although both polynomiality and positivity are basic properties that must
hold in any reasonable model of GPDs usually the model building community
meets a dilemma: one can use the double distribution representation (\ref%
{DD-representation}) but it does not guarantee that the infinite set of
inequalities (\ref{ineq-q}) will be satisfied \cite{MMPR-02,TM-02}.
Alternatively one can build the models based on the representation (\ref%
{pos-representation}) or on the so called overlap representation \cite%
{DFJK-00} which also automatically guarantees positivity bounds but then one
meets problems with the polynomiality. In this paper a representation for
GPDs is suggested which guarantees both positivity and polynomiality.

\section{Modified double distribution representation}

\label{modified-DD-section}

For the construction of GPDs $H^{(N)}(x,\xi ,t)$ obeying both polynomiality
and positivity constraints we use the double distribution representation
which differs from the standard representation (\ref{DD-representation}) by the extra
factor of $(1-x)^{N}$%
\[
H^{(N)}(x,\xi ,t)=(1-x)^{N}
\]
\begin{equation}
\times \int\limits_{|\alpha |+|\beta |\leq 1}d\alpha d\beta \delta (x-\xi
\alpha -\beta )F_{D}^{\prime }(\alpha ,\beta ,t)\,.  \label{F-N-new}
\end{equation}%
Here $N$ depends on the type of the parton distribution according to Table 
\ref{table-N}.

Representation (\ref{F-N-new}) obviously satisfies the polynomiality
condition (\ref{polynomiality-power}). Indeed,%
\[
\int\limits_{-1}^{1}dx\,x^{n}\int\limits_{|\alpha |+|\beta |\leq 1}d\alpha
d\beta \delta (x-\xi \alpha -\beta )F_{D}^{\prime }(\alpha ,\beta ,t)
\]
\begin{equation}
=P_{n}(\xi ,t)
\end{equation}%
where $P_{n}(\xi ,t)$ is a polynomial of degree $n.$ Therefore%
\[
\int\limits_{-1}^{1}dx\,x^{m}H^{(N)}(x,\xi
,t)=\int\limits_{-1}^{1}dx\,\int\limits_{|\alpha |+|\beta |\leq 1}d\alpha
d\beta x^{m}(1-x)^{N}
\]
\begin{equation}
\times \delta (x-\xi \alpha -\beta )F_{D}^{\prime }(\alpha ,\beta
,t)=S_{N+m}(\xi ,t)
\end{equation}%
is a polynomial in $\xi $ of degree $N+m$ in agreement with
Eq. (\ref{polynomiality-power}).

For our aims it is convenient to use parameters 
\begin{equation}
\alpha _{1}=\frac{1}{2}(1-\beta -\alpha )\,,\quad \alpha _{2}=\frac{1}{2}%
(1-\beta +\alpha ).
\end{equation}%
instead of $\alpha ,\beta $. Actually it is $\alpha _{1}$ and $\alpha _{2}$ (%
$\alpha _{1}=1-x-y$, $\alpha _{2}=y$ in terms of variables $x,y$ used in
refs. \cite{Radyushkin-96-a,Radyushkin-97}) that appear as $\alpha $
parameters in the perturbative diagrammatic justification of the double
distribution representation. The modified double distribution expressed in
terms of parameters $\alpha _{1},\alpha _{2}$ will be denoted as follows:%
\begin{equation}
F_{D}(\alpha _{1},\alpha _{2},t)\equiv F_{D}^{\prime }(\alpha ,\beta ,t)\,.
\end{equation}%
After these changes the modified double distribution representation (\ref%
{F-N-new}) takes the following form:
\[
H^{(N)}\left( x,\xi ,t\right) =2(1-x)^{N}\int\limits_{0}^{1}d\alpha
_{1}\int\limits_{0}^{1-\alpha _{1}}d\alpha _{2}
\]
\begin{equation}
\times F_{D}\left( \alpha _{1},\alpha _{2},t\right) \delta \left[ x-\xi
(\alpha _{2}-\alpha _{1})-(1-\alpha _{1}-\alpha _{2})\right] \,.
\label{GPD-DD}
\end{equation}%
Here we use the triangle integration region in the $\alpha _{1},\alpha _{2}$
plane which corresponds to the constraint $\beta >0$ in terms of variables $%
\alpha ,\beta $. Hence our GPD vanishes in the ``antiquark'' region (for
brevity we use the word ``quark'' for any type of partons):%
\begin{equation}
H^{(N)}\left( x,\xi ,t\right) =0\quad \mathrm{if}\quad x<-|\xi |\,.
\end{equation}%
Therefore we must take care about the positivity constraints only in the
``quark'' region $x>|\xi |$. Once this pure quark GPD is constructed we can
use the transformation $x\rightarrow -x$ to build GPDs with appropriate
properties in both quark and antiquark regions.

In the case $N>0$ one can add the $D$-term to the modified double
distribution representation (\ref{GPD-DD}):
\[
H^{(N)}\left( x,\xi ,t\right) \stackrel{N>0}{\rightarrow }H^{(N)}\left( x,\xi
,t\right)
\]
\begin{equation}
+x^{N-1}D\left( \frac{x}{\xi },t\right) \theta \left( 1-\left| \frac{x}{\xi }%
\right| \right) \mathrm{sign}(\xi )\,.  \label{F-D-terms}
\end{equation}%
In principle, using the trick of Ref.~\cite{BMKS-01}, one can include the $D$%
-term into the double distribution. But if one is interested in a
parametrization of GPDs obeying the positivity and polynomiality constraints,
then the $D$-term is useful: it is localized in the region $|x|<|\xi |$ and
therefore it does not appear in the positivity condition (\ref{ineq-q}). The polynomiality is obvious for the $D$-term. Thus by
adding an arbitrary $D$-term we violate neither polynomiality nor positivity.

\section{Ansatz for double distributions}

\label{DD-ansatz-section}

Now the problem is to find double distributions $F_{D}\left( \alpha
_{1},\alpha _{2},t\right) $ which lead to GPDs $H^{(N)}\left( x,\xi
,t\right) $ obeying the positivity constraint. We use the following ansatz
for the modified double distributions (\ref{GPD-DD}):
\[
F_{D}\left( \alpha _{1},\alpha _{2},t\right) =\int\limits_{0}^{\infty
}d\lambda \,\lambda \int d\nu
\]

\begin{equation}
\times \left( \frac{1}{\lambda \alpha _{1}\alpha _{2}}-t\right) ^{-\nu
-1}L_{\nu }(\lambda \alpha _{1},\lambda \alpha _{2})\,.  \label{DD-ansatz}
\end{equation}%
Our double distribution is parametrized by an infinite set of functions $%
L_{\nu }(w_{1},w_{2})$ defined for $w_{1},w_{2}\geq 0$ and depending on
parameter $\nu $. We assume that for any $\nu $ function $L_{\nu
}(w_{1},w_{2})$ corresponds to a positive definite quadratic form in $%
w_{1},w_{2}$, i.e. for any function $\phi (w)$%
\begin{equation}
\int\limits_{0}^{\infty }dw_{1}\int\limits_{0}^{\infty }dw_{2}L_{\nu
}(w_{1},w_{2})\phi (w_{1})\phi ^{\ast }(w_{2})\geq 0\,.
\label{L-K-positivie}
\end{equation}%
This is equivalent to the existence of the following integral representation
for $L_{\nu }(w_{1},w_{2})$

\begin{equation}
L_{\nu }(w_{1},w_{2})=\int d\rho F_{\nu }\,(w_{1},\rho )F_{\nu }^{\ast
}\left( w_{2},\rho \right)  \label{L-F-decompos}
\end{equation}%
or to its discrete series analog.
Since we are interested in real and $\xi$-even GPDs, we must use real
functions $L_\nu$ and $F_\nu$.

The lower limit of the integral over $\nu $ on the right-hand side (RHS) of
Eq. (\ref{DD-ansatz}) determines the asymptotics of $F_{D}\left( \alpha
_{1},\alpha _{2},t\right) $ at large $|t|$. If one integrates over positive $%
\nu $, then $F_{D}\sim |t|^{-1}$. Functions $L_{\nu }$ appearing in Eq. (\ref%
{DD-ansatz}) have the $\nu $ dependent dimension, which is slightly awkward
but simplifies the equations.

Since $\lambda ,\alpha _{1},\alpha _{2}\geq 0$ and $t\leq 0$ the following
factor appearing in Eq. (\ref{DD-ansatz}) is always positive:
\begin{equation}
\frac{1}{\lambda \alpha _{1}\alpha _{2}}-t>0\,.
\end{equation}

Below it will be shown that for any set of positive definite functions $%
L_{\nu }(w_{1},w_{2})$ under the assumption that the integrals on the RHS of
(\ref{DD-ansatz}) are convergent, the resulting double distribution $%
F_{D}\left( \alpha _{1},\alpha _{2},t\right) $ (\ref{DD-ansatz}) leads to
the GPD $H^{(N)}\left( x,\xi ,t\right) $ (\ref{GPD-DD}) which satisfies the
positivity bound (\ref{ineq-q}). This check of positivity will be done in
Section \ref{positivity-check-section} but first we prefer to derive some
useful relations.

\section{Expression for GPDs}

\label{GPDs-section}

Let us derive the expressions for GPDs $H^{(N)}\left( x,\xi ,t\right) $
corresponding to the double distribution (\ref{DD-ansatz}). First we insert
ansatz (\ref{DD-ansatz}) for the double distribution $F_{D}\left( \alpha
_{1},\alpha _{2},t\right) $ into representation (\ref{GPD-DD}) for GPD $%
H^{(N)}\left( x,\xi ,t\right) $
\[
H^{(N)}\left( x,\xi ,t\right) =2(1-x)^{N}\int\limits_{0}^{1}d\alpha
_{1}\int\limits_{0}^{1-\alpha _{1}}d\alpha _{2}
\]
\[
\times \delta \left[ x-\xi (\alpha _{2}-\alpha _{1})-(1-\alpha _{1}-\alpha
_{2})\right]
\]
\begin{equation}
\times \int\limits_{0}^{\infty }d\lambda \,\lambda \int d\nu \left( \frac{1}{%
\lambda \alpha _{1}\alpha _{2}}-t\right) ^{-\nu -1}L_{\nu }(\lambda \alpha
_{1},\lambda \alpha _{2})\,.  \label{F-L-K-gamma}
\end{equation}%
We can rewrite this as follows:
\[
H^{(N)}\left( x,\xi ,t\right) =2(1-x)^{N}\int\limits_{0}^{\infty }d\lambda
\,\lambda \int\limits_{0}^{\infty }d\alpha _{1}\int\limits_{0}^{\infty
}d\alpha _{2}
\]
\[
\times \theta (1-\alpha _{1}-\alpha _{2})\delta \left[ x-\xi (\alpha
_{2}-\alpha _{1})-(1-\alpha _{1}-\alpha _{2})\right]
\]
\begin{equation}
\times \int d\nu \left( \frac{1}{\lambda \alpha _{1}\alpha _{2}}-t\right)
^{-\nu -1}L_{\nu }(\lambda \alpha _{1},\lambda \alpha _{2})\,.
\end{equation}%
Let us introduce new integration variables%
\begin{equation}
w_{k}=\lambda \alpha _{k}
\end{equation}%
instead of $\alpha _{k}$ and integrate over $\lambda $ using the delta
function%
\[
H^{(N)}\left( x,\xi ,t\right) =2(1-x)^{N-1}
\]
\[
\times \int\limits_{0}^{\infty }dw_{1}\int\limits_{0}^{\infty }dw_{2}\theta
\left( x-\xi \frac{w_{2}-w_{1}}{w_{1}+w_{2}}\right)
\]
\begin{equation}
\times \int d\nu \left( \frac{1}{w_{2}}\frac{1+\xi }{1-x}+\frac{1}{w_{1}}%
\frac{1-\xi }{1-x}-t\right) ^{-\nu -1}L_{\nu }(w_{1},w_{2})\,.  \label{F-GPD}
\end{equation}

The step function does not vanish in the region $x>|\xi |$ so that
the above expression simplifies as follows:
\[
\left. H^{(N)}\left( x,\xi ,t\right) \right| _{x>|\xi
|}=2(1-x)^{N-1}\int\limits_{0}^{\infty }dw_{1}\int\limits_{0}^{\infty }dw_{2}
\]
\begin{equation}
\times \int d\nu \left( \frac{1}{w_{2}}\frac{1+\xi }{1-x}+\frac{1}{w_{1}}%
\frac{1-\xi }{1-x}-t\right) ^{-\nu -1}L_{\nu }(w_{1},w_{2})\,.
\label{F-GPD-DGLAP}
\end{equation}%
This representation can be rewritten in the following form:
\[
\left. H^{(N)}\left( x,\xi ,t\right) \right| _{x>|\xi
|}=2(1-x)^{N-1}\int\limits_{0}^{\infty }dw_{1}\int\limits_{0}^{\infty }dw_{2}
\]
\[
\times \int\limits_{0}^{\infty }d\gamma e^{t\gamma }\int d\nu \frac{\gamma
^{\nu }}{\Gamma (\nu +1)}L_{\nu }(w_{1},w_{2})
\]
\begin{equation}
\times \exp \left[ -\gamma \left( \frac{1}{w_{2}}\frac{1+\xi }{1-x}+\frac{1}{%
w_{1}}\frac{1-\xi }{1-x}\right) \right] \,.  \label{F-L-K-1}
\end{equation}

\section{Forward distribution}

\label{forward-section}

In the forward limit $\xi \rightarrow 0,$ $t\rightarrow 0$ we obtain from
Eq. (\ref{F-GPD-DGLAP}) 
\[
f(x)=H^{(N)}\left( x,0,0\right) =2(1-x)^{N-1}
\]
\begin{equation}
\times \int\limits_{0}^{\infty }dw_{1}\int\limits_{0}^{\infty }dw_{2}\int
d\nu \left( \frac{w_{1}w_{2}(1-x)}{w_{1}+w_{2}}\right) ^{\nu +1}L_{\nu
}(w_{1},w_{2})\,.  \label{q-forward}
\end{equation}%
The positivity of forward parton distributions is a consequence of the
general positivity bounds on GPDs which will be established in the next
section. On the other hand, we can see the positivity of the forward parton
distribution $f(x)$ directly from Eq. (\ref{q-forward}): 
\[
\int\limits_{0}^{\infty }dw_{1}\int\limits_{0}^{\infty }dw_{2}\left( \frac{%
w_{1}w_{2}}{w_{1}+w_{2}}\right) ^{\nu +1}L_{\nu }(w_{1},w_{2})
\]
\[
=\int\limits_{0}^{\infty }dw_{1}\int\limits_{0}^{\infty }dw_{2}\frac{1}{%
\Gamma (\nu +1)}\int\limits_{0}^{\infty }d\tau \,\tau ^{\nu }
\]
\[
\times \exp \left( -\tau \frac{w_{1}+w_{2}}{w_{1}w_{2}}\right) L_{\nu
}(w_{1},w_{2})
\]
\[
=\frac{1}{\Gamma (\nu +1)}\int\limits_{0}^{\infty }d\tau \,\tau ^{\nu
}\int\limits_{0}^{\infty }dw_{1}\int\limits_{0}^{\infty }dw_{2}
\]
\begin{equation}
\times L_{\nu }(w_{1},w_{2})\exp \left( -\frac{\tau }{w_{1}}\right) \exp
\left( -\frac{\tau }{w_{2}}\right) \geq 0\,.
\end{equation}%
The positivity of the RHS follows from the inequality 
\begin{equation}
\int\limits_{0}^{\infty }dw_{1}\int\limits_{0}^{\infty }dw_{2}L_{\nu
}(w_{1},w_{2})\exp \left( -\frac{\tau }{w_{1}}\right) \exp \left( -\frac{%
\tau }{w_{2}}\right) \geq 0\,
\end{equation}%
which is a consequence of the positivity (\ref{L-K-positivie}) of the
quadratic form $L_{\nu }(w_{1},w_{2})$.

\section{Proof of positivity}

\label{positivity-check-section}

Now we want to show that the modified double distribution (\ref{DD-ansatz})
with positive definite functions $L_{\nu }$ (\ref{L-K-positivie}) generates
GPD $H^{(N)}\left( x,\xi ,t\right) $ which satisfies the positivity bounds (%
\ref{ineq-q}). For the positivity bounds we need the GPD in the impact
parameter representation (\ref{F-impact-def})%
\[
\tilde{F}^{(N)}\left( x,\xi ,\frac{1-x}{1-\xi ^{2}}b^{\perp }\right) =\int 
\frac{d^{2}\Delta ^{\perp }}{(2\pi )^{2}}\exp \left[ i\frac{1-x}{1-\xi ^{2}}%
(\Delta ^{\perp }b^{\perp })\right]
\]
\begin{equation}
\times H^{(N)}\left( x,\xi ,-\frac{|\Delta ^{\perp }|^{2}+4\xi ^{2}M^{2}}{%
1-\xi ^{2}}\right) \,.
\end{equation}%
Using representation (\ref{F-L-K-1}) for the GPDs $H^{(N)}\left( x,\xi
,t\right)$, we obtain

\[
\tilde{F}^{(N)}\left( x,\xi ,\frac{1-x}{1-\xi ^{2}}b^{\perp }\right) =\int 
\frac{d^{2}\Delta ^{\perp }}{(2\pi )^{2}}\exp \left[ i\frac{1-x}{1-\xi ^{2}}%
(\Delta ^{\perp }b^{\perp })\right]
\]
\[
\times 2(1-x)^{N-1}\int\limits_{0}^{\infty }dw_{1}\int\limits_{0}^{\infty
}dw_{2}\int\limits_{0}^{\infty }d\gamma
\]
\[
\times \exp \left( -\gamma \frac{|\Delta ^{\perp }|^{2}+4\xi ^{2}M^{2}}{%
1-\xi ^{2}}\right) \int d\nu \frac{\gamma ^{\nu }}{\Gamma (\nu +1)}
\]
\begin{equation}
\times \exp \left[ -\gamma \left( \frac{1}{w_{2}}\frac{1+\xi }{1-x}+\frac{1}{%
w_{1}}\frac{1-\xi }{1-x}\right) \right] L_{\nu }\left( w_{1},w_{2}\right) \,.
\end{equation}%
Integrating over $\Delta ^{\perp }$, introducing compact notation
\begin{equation}
r_{1}=\frac{1-x}{1+\xi },\quad r_{2}=\frac{1-x}{1-\xi }\,.
\end{equation}%
(see Appendix \ref{r-k-appendix}), and rescaling the integration variables $%
w_{k}\rightarrow w_{k}r_{k}$, we find%
\[
\tilde{F}^{(N)}\left( x,\xi ,\frac{1-x}{1-\xi ^{2}}b^{\perp }\right) =\frac{1%
}{2\pi }\left( \frac{2r_{1}r_{2}}{r_{1}+r_{2}}\right)
^{N+1}\int\limits_{0}^{\infty }\frac{d\gamma }{\gamma }
\]
\[
\times \int\limits_{0}^{\infty }dw_{1}\int\limits_{0}^{\infty }dw_{2}\exp 
\left[ -\frac{r_{1}r_{2}}{4\gamma }|b^{\perp }|^{2}-\gamma \frac{%
M^{2}(r_{1}-r_{2})^{2}}{r_{1}r_{2}}\right]
\]
\begin{equation}
\times \int d\nu \frac{\gamma ^{\nu }}{\Gamma (\nu +1)}\exp \left[ -\frac{%
\gamma (w_{1}+w_{2})}{r_{1}r_{2}w_{1}w_{2}}\right] L_{\nu }\left(
r_{1}w_{1},r_{2}w_{2}\right) \,.
\end{equation}%
Next we change the integration variable $\gamma \rightarrow \gamma
r_{1}r_{2} $%
\[
\tilde{F}^{(N)}\left( x,\xi ,\frac{1-x}{1-\xi ^{2}}b^{\perp }\right) =\frac{1%
}{2\pi }\left( \frac{2r_{1}r_{2}}{r_{1}+r_{2}}\right)
^{N+1}\int\limits_{0}^{\infty }\frac{d\gamma }{\gamma }
\]
\begin{equation}
\times \int\limits_{0}^{\infty }dw_{1}\int\limits_{0}^{\infty }dw_{2}\exp 
\left[ -\frac{1}{4\gamma }|b^{\perp }|^{2}-\gamma M^{2}(r_{1}-r_{2})^{2}%
\right]
\end{equation}%
\begin{equation}
\times \int d\nu \frac{(\gamma r_{1}r_{2})^{\nu }}{\Gamma (\nu +1)}\exp %
\left[ -\frac{\gamma (w_{1}+w_{2})}{w_{1}w_{2}}\right] L_{\nu }\left(
r_{1}w_{1},r_{2}w_{2}\right)
\end{equation}%
and use the representation%
\[
\exp \left[ -\gamma (r_{1}-r_{2})^{2}M^{2}\right]
\]
\begin{equation}
=\frac{1}{2M\sqrt{\pi \gamma }}\int\limits_{-\infty }^{\infty }ds\exp \left[
-\frac{s^{2}}{4\gamma M^{2}}+is(r_{2}-r_{1})\right] \,.
\end{equation}%
Then%
\[
\tilde{F}^{(N)}\left( x,\xi ,\frac{1-x}{1-\xi ^{2}}b^{\perp }\right) =\frac{1%
}{4M\pi ^{3/2}}\left( \frac{2r_{1}r_{2}}{r_{1}+r_{2}}\right) ^{N+1}
\]
\[
\times \int\limits_{0}^{\infty }d\gamma \exp \left( -\frac{1}{4\gamma }%
|b^{\perp }|^{2}\right) \int\limits_{-\infty }^{\infty }ds\exp \left( -\frac{%
s^{2}}{4\gamma M^{2}}\right)
\]
\[
\times \int\limits_{0}^{\infty }dw_{1}\int\limits_{0}^{\infty }dw_{2}\exp 
\left[ is(r_{2}-r_{1})-\frac{\gamma (w_{1}+w_{2})}{w_{1}w_{2}}\right]
\]
\begin{equation}
\times \int d\nu \gamma ^{\nu -3/2}\frac{(r_{1}r_{2})^{\nu }}{\Gamma (\nu +1)%
}L_{\nu }\left( w_{1}r_{1},w_{2}r_{2}\right) \,.
\end{equation}%
Now we turn to the positivity bound (\ref{ineq-q}) written in the form of
the integral over $r_{1},r_{2}$ --- see equation (\ref{uv-ineq-2}) in
Appendix \ref{r-k-appendix}. The left-hand side of this inequality is

\[
\,\int\limits_{0}^{1}dr_{1}\int\limits_{0}^{1}dr_{2}\left(
r_{1}+r_{2}\right) ^{N+1}p^{\ast }\left( r_{2}\right) p\left( r_{1}\right)
\]
\[
\times \tilde{F}^{(N)}\left( x,\xi ,\frac{1-x}{1-\xi ^{2}}b^{\perp }\right)
\]
\[
=\frac{2^{N-1}}{M\pi ^{3/2}}\int d\nu \int\limits_{0}^{\infty }d\gamma \frac{%
\gamma ^{\nu -3/2}}{\Gamma (\nu +1)}\exp \left( -\frac{1}{4\gamma }|b^{\perp
}|^{2}\right)
\]
\[
\times \int\limits_{-\infty }^{\infty }ds\exp \left( -\frac{s^{2}}{4\gamma
M^{2}}\right) \int\limits_{0}^{1}dr_{1}\int\limits_{0}^{\infty
}dw_{1}\int\limits_{0}^{1}dr_{2}\int\limits_{0}^{\infty }dw_{2}
\]
\[
\times L_{\nu }\left( w_{1}r_{1},w_{2}r_{2}\right) \left[ p\left(
r_{1}\right) r_{1}^{N+\nu +1}\exp \left( -isr_{1}-\frac{\gamma }{w_{1}}%
\right) \right]
\]
\begin{equation}
\times \left[ p\left( r_{2}\right) r_{2}^{N+\nu +1}\exp \left( -isr_{2}-%
\frac{\gamma }{w_{2}}\right) \right] ^{\ast }\,.  \label{ineq-lhs}
\end{equation}%
Here we can rescale integration variables $w_{k}\rightarrow w_{k}/r_{k}$.
Then%
\[
\int\limits_{0}^{1}dr_{1}\int\limits_{0}^{\infty
}dw_{1}\int\limits_{0}^{1}dr_{2}\int\limits_{0}^{\infty }dw_{2}L_{\nu
}\left( w_{1}r_{1},w_{2}r_{2}\right)
\]
\[
\times \left[ p\left( r_{1}\right) r_{1}^{N+\nu +1}\exp \left( -isr_{1}-%
\frac{\gamma }{w_{1}}\right) \right]
\]
\[
\times \left[ p\left( r_{2}\right) r_{2}^{N+\nu +1}\exp \left( -isr_{2}-%
\frac{\gamma }{w_{2}}\right) \right] ^{\ast }
\]
\[
=\int\limits_{0}^{\infty }dw_{1}\int\limits_{0}^{\infty }dw_{2}L_{\nu
}\left( w_{1},w_{2}\right)
\]
\[
\times \left[ \int\limits_{0}^{1}dr_{1}p\left( r_{1}\right) r_{1}^{N+\nu
}\exp \left( -isr_{1}-\frac{\gamma r_{1}}{w_{1}}\right) \right]
\]
\[
\times \left[ \int\limits_{0}^{1}dr_{2}p\left( r_{2}\right) r_{2}^{N+\nu
}\exp \left( -isr_{2}-\frac{\gamma r_{2}}{w_{2}}\right) \right] ^{\ast }
\]
\begin{equation}
=\int\limits_{0}^{\infty }dw_{1}\int\limits_{0}^{\infty }dw_{2}L_{\nu
}\left( w_{1},w_{2}\right) \phi _{\nu }(w_{1})\phi _{\nu }^{\ast
}(w_{2})\geq 0  \label{last-ineq}
\end{equation}%
where%
\begin{equation}
\phi _{\nu }(w)=\int\limits_{0}^{1}drp(r)r^{N+\nu }\exp \left( -isr-\frac{%
\gamma r}{w}\right) \,.
\end{equation}%
The RHS of Eq. (\ref{last-ineq}) is positive since $L_{\nu }\left(
w_{1},w_{2}\right) $ is positive definite. Combining Eqs. (\ref{ineq-lhs})
and (\ref{last-ineq}), we complete the proof of the positivity bound (\ref%
{uv-ineq-2}) for the GPD generated by the double distribution (\ref%
{DD-ansatz}).

\section{Conclusions}

In this paper we have shown that representation (\ref{DD-ansatz}) for the
double distributions [understood in the sense of Eq. (\ref{GPD-DD})]
generates GPDs (\ref{F-GPD}) satisfying both polynomiality and positivity
constraints. Our representation (\ref{DD-ansatz}) for double distributions
involves arbitrary positive definite quadratic forms $L_{\nu }(w_{1},w_{2})$%
. Functions $L_{\nu }(w_{1},w_{2})$ parametrizing GPDs depend on the same
amount of variables ($w_{1},w_{2},\nu $) as GPDs themselves ($x,\xi ,t$).
This means that the class of solutions of the polynomiality and positivity
constraints found in this paper is rather wide. On the other hand, this set
of solutions is not complete. Indeed, our ansatz (\ref{DD-ansatz}) does not
depend on the mass of the hadron $M$ whereas the positivity and
polynomiality constraints are sensitive to $M$: although $M$ appears neither
in the polynomiality condition (\ref{polynomiality-power}) for $H^{(N)}$ nor
in the positivity bound (\ref{ineq-q}) for $\tilde{F}^{(N)}$, the relation (%
\ref{F-impact-def}) between $H^{(N)}$ and $\tilde{F}^{(N)}$ contains the
hadron mass $M$. This means that the combined constraints of positivity and
polynomiality are sensitive to the hadron mass $M$. Therefore the absence of
the $M$ dependence in our ansatz (\ref{DD-ansatz}) should mean that there
must exist other solutions of the polynomiality and positivity constraints
and one has to try other methods in order to find the other solutions. In
particular, in Ref. \cite{Pobylitsa-02-d} the solutions of the positivity
and polynomiality constraints are constructed in terms of triangle
perturbative diagrams.

The parametrization of GPDs suggested here seems to be constructive for the
model building: the positive definite functions $L_{\nu }(w_{1},w_{2})$ can
be easily generated by using Eq. (\ref{L-F-decompos}). One should not forget
about the possibility to add the $D$-term (\ref{F-D-terms}) which is not
constrained by the polynomiality and positivity.

Certainly apart from the positivity and polynomiality there are other
theoretical and phenomenological constraints on GPDs and it would be
interesting whether representation (\ref{DD-ansatz}) allows to construct
viable models of GPDs.

\textbf{Acknowledgments.} I am grateful to A.V.~Belitsky, M.~Diehl,
R. Jakob, P.~Kroll, D.~M\"uller, M.V.~Polyakov and A.V.~Radyushkin for
useful discussions. This work was supported by DFG and BMBF.

\appendix

\section{Solution of the positivity bounds}

\label{r-k-appendix}

In this appendix we derive the solution (\ref{pos-representation}) of the
positivity bounds (\ref{ineq-q}). First let us define variables $r_{1},r_{2}$
which can be used instead of $x,\xi $%
\begin{equation}
r_{1}=\frac{1-x}{1+\xi },\quad r_{2}=\frac{1-x}{1-\xi }\,,  \label{r-12-def}
\end{equation}%
\begin{equation}
\xi =\frac{r_{2}-r_{1}}{r_{2}+r_{1}}\,,\quad x=1-\frac{2r_{1}r_{2}}{%
r_{1}+r_{2}}\,,
\end{equation}%
\begin{equation}
\frac{2dxd\xi }{(1-x)^{3}}=\frac{dr_{1}dr_{2}}{r_{1}^{2}r_{2}^{2}}\,.
\label{u-v-measure}
\end{equation}%
The region covered by the positivity bounds (\ref{ineq-q})%
\begin{equation}
x>|\xi |
\end{equation}%
is mapped to the square in the $r_{1},r_{2}$ plane
\begin{equation}
0<r_{1},r_{2}<1\,.
\end{equation}%
Inequality (\ref{ineq-q}) takes the following form in terms of integration
variables $r_{1},r_{2}$ (we keep variables $x,\xi $ in GPDs implying that
they are functions of $r_{1},r_{2}$):
\[
\,\int\limits_{0}^{1}\frac{dr_{1}}{r_{1}^{2}}\int\limits_{0}^{1}\frac{dr_{2}%
}{r_{2}^{2}}\left( \frac{r_{1}+r_{2}}{r_{1}r_{2}}\right) ^{N+1}p^{\ast
}\left( r_{2}\right) p\left( r_{1}\right)
\]
\begin{equation}
\times \tilde{F}^{(N)}\left( x,\xi ,\frac{1-x}{1-\xi ^{2}}b^{\perp }\right)
\geq 0\,\,.  \label{uv-ineq-1}
\end{equation}%
Since function $p$ is arbitrary we can replace it%
\begin{equation}
p(r_{1})\rightarrow r_{1}^{N+3}p(r_{1})
\end{equation}%
which leads us to the equivalent form of inequality (\ref{uv-ineq-1})%
\[
\int\limits_{0}^{1}dr_{1}\int\limits_{0}^{1}dr_{2}\left(
r_{1}+r_{2}\right) ^{N+1}
\]
\begin{equation}
\times p^{\ast }\left( r_{2}\right) p\left( r_{1}\right) \tilde{F}%
^{(N)}\left( x,\xi ,\frac{1-x}{1-\xi ^{2}}b^{\perp }\right) \geq 0\,\,.
\label{uv-ineq-2}
\end{equation}%
Inequality (\ref{uv-ineq-1}) means that function%
\begin{equation}
\left( \frac{r_{1}+r_{2}}{r_{1}r_{2}}\right) ^{N+1}\tilde{F}^{(N)}\left(
x,\xi ,\frac{1-x}{1-\xi ^{2}}b^{\perp }\right)
\end{equation}%
must be a positive definite quadratic form, i.e. it has the following
representation%
\[
\left( \frac{r_{1}+r_{2}}{2r_{1}r_{2}}\right) ^{N+1}\tilde{F}^{(N)}\left(
x,\xi ,\frac{1-x}{1-\xi ^{2}}b^{\perp }\right)
\]
\begin{equation}
=\sum\limits_{n}R_{n}(r_{1},b^{\perp })R_{n}^{\ast }(r_{2},b^{\perp })
\end{equation}%
with some functions $R_{n}$. Turning back to the variables $x,\xi$, we find
\[
\tilde{F}^{(N)}\left( x,\xi ,b^{\perp }\right) =(1-x)^{N+1}
\]
\begin{equation}
\times \sum\limits_{n}R_{n}\left( \frac{1-x}{1+\xi },\frac{1-\xi ^{2}}{1-x}%
b^{\perp }\right) R_{n}^{\ast }\left( \frac{1-x}{1-\xi },\frac{1-\xi ^{2}}{%
1-x}b^{\perp }\right) \,.  \label{F-Q-deriv}
\end{equation}
In the case of real and $\xi$-even GPDs, functions $R_n$ are real.

Introducing functions
\begin{equation}
Q_{n}(r,b^{\perp })=R_{n}\left( r,\frac{1}{r}b^{\perp }\right)
\end{equation}%
we obtain representation (\ref{pos-representation}) for $\tilde{F}%
^{(N)}\left( x,\xi ,b^{\perp }\right) $.

\end{document}